\documentstyle[prd,aps,preprint,tighten,epsfig]{revtex}

\begin{document}

\draft

\title{Radiative Generation of Leptonic CP Violation}
\author{{\bf Shu Luo} ~ and ~ {\bf Jianwei Mei}}
\address{Institute of High Energy Physics, Chinese Academy of
Sciences, \\
P.O. Box 918, Beijing 100049, China}
\author{\bf Zhi-zhong Xing}
\address{CCAST (World Laboratory), P.O. Box 8730, Beijing 100080,
China \\
and Institute of High Energy Physics, Chinese Academy of Sciences,
\\
P.O. Box 918, Beijing 100049, China \footnote{Mailing address}
\\
({\it Electronic address: xingzz@mail.ihep.ac.cn})}
\maketitle

\begin{abstract}
Three CP-violating phases of the $3\times 3$ lepton flavor mixing
matrix $V$ are entangled with one another in the
renormalization-group evolution from the seesaw scale
($\Lambda_{\rm SS} \sim 10^{14}$ GeV) to the electroweak scale
($\Lambda_{\rm EW} \sim 10^2$ GeV). Concerning the Dirac phase
$\delta$, we show that $\delta =90^\circ$ at $\Lambda_{\rm EW}$
can be radiatively generated from $\delta =0^\circ$ at
$\Lambda_{\rm SS}$ in the minimal supersymmetric standard model,
if three neutrino masses are nearly degenerate. As for the
Majorana phases $\rho$ and $\sigma$, it is also possible to
radiatively generate $\rho =90^\circ$ or $\sigma = 90^\circ$ at
$\Lambda_{\rm EW}$ from $\rho =0^\circ$ or $\sigma = 0^\circ$ at
$\Lambda_{\rm SS}$. The one-loop renormalization-group equations
for the Jarlskog invariant and two off-diagonal asymmetries of $V$
are derived, and their running behaviors from $\Lambda_{\rm SS}$
to $\Lambda_{\rm EW}$ are numerically illustrated.
\end{abstract}

\pacs{PACS number(s): 14.60.Pq, 13.10.+q, 25.30.Pt}

\newpage

\section{Introduction}

Recent solar \cite{SNO}, atmospheric \cite{SK}, reactor (KamLAND
\cite{KM} and CHOOZ \cite{CHOOZ}) and accelerator (K2K \cite{K2K})
neutrino oscillation experiments have provided us with very robust
evidence that neutrinos are massive and lepton flavors are mixed.
The phenomenon of lepton flavor mixing can be described by a
$3\times 3$ unitary matrix $V$, commonly referred to as the
Maki-Nakagawa-Sakata (MNS) matrix \cite{MNS}. A useful
parametrization of $V$ reads \cite{FX01}:
\begin{equation}
V = \left( \matrix{ c^{}_{12}c^{}_{13} & s^{}_{12}c^{}_{13} &
s^{}_{13} \cr -c^{}_{12}s^{}_{23}s^{}_{13} - s^{}_{12}c^{}_{23}
e^{-i\delta} & -s^{}_{12}s^{}_{23}s^{}_{13} + c^{}_{12}c^{}_{23}
e^{-i\delta} & s^{}_{23}c^{}_{13} \cr -c^{}_{12}c^{}_{23}s^{}_{13}
+ s^{}_{12}s^{}_{23} e^{-i\delta} & -s^{}_{12}c^{}_{23}s^{}_{13} -
c^{}_{12}s^{}_{23} e^{-i\delta} & c^{}_{23}c^{}_{13} } \right)
\left ( \matrix{e^{i\rho } & 0 & 0 \cr 0 & e^{i\sigma} & 0 \cr 0 &
0 & 1 \cr} \right ) \; ,
%       (1)
\end{equation}
where $c^{}_{ij} \equiv \cos\theta_{ij}$ and $s^{}_{ij} \equiv
\sin\theta_{ij}$ (for $ij=12,23$ and $13$). Note that the
CP-violating phase $\delta$ governs the strength of CP or T
violation in normal neutrino oscillations, and it does not appear
in the expression of $\langle m\rangle_{ee}$
--- the effective mass term of the neutrinoless double-beta decay
%%%%%%%%%%%%%%%%%%%%%%%%%%%%%
\footnote{Namely, $\langle m\rangle_{ee} = \left | m^{}_1 c^2_{12}
c^2_{13} e^{2i\rho} + m^{}_2 s^2_{12} c^2_{13} e^{2i\sigma} +
m^{}_3 s^2_{13} \right |$ is independent of $\delta$.}.
%%%%%%%%%%%%%%%%%%%%%%%%%%%%%
On the other hand, the CP-violating phases $\rho$ and $\sigma$
only affect $\langle m\rangle_{ee}$, and they can be rotated away
if the massive neutrinos are Dirac particles. Current experimental
data indicate $\theta_{12} \approx 33^\circ$, $\theta_{23} \approx
45^\circ$ and $\theta_{13} < 10^\circ$ \cite{Fit}, but the phase
parameters $\delta$, $\rho$ and $\sigma$ are entirely
unrestricted. A variety of new neutrino experiments are underway,
not only to measure the smallest mixing angle $\theta_{13}$ and
the Dirac phase $\delta$, but also to constrain the Majorana
phases $\rho$ and $\sigma$.

While neutrino masses and lepton flavor mixing parameters can be
measured at low-energy scales, their origin is most likely to
depend on some unspecified interactions at a superhigh energy
scale. For instance, the existence of very heavy right-handed
neutrinos and lepton number violation may naturally explain the
smallness of left-handed neutrino masses via the famous seesaw
mechanism \cite{SS} at the scale $\Lambda_{\rm SS} \sim 10^{14}$
GeV. Below this seesaw scale, the effective Lagrangian for lepton
Yukawa interactions can be written as
\begin{equation}
-{\cal L} = \overline{E_L^{}}H_1 Y_l^{} l_R^{} - \frac{1}{2}
\overline{E_L^{}} H_2 \cdot\kappa\cdot H_2^{c\dag} E_L^c ~ + ~
{\rm h.c.} \;
%       (2)
\end{equation}
in the minimal supersymmetric standard model (MSSM)
%%%%%%%%%%%%%%%%%%%%
\footnote{For the sake of simplicity, we assume the supersymmetry
breaking scale $\Lambda_{\rm SUSY}$ to be close to the electroweak
scale $\Lambda_{\rm EW}$. Even if $\Lambda_{\rm SUSY}/\Lambda_{\rm
EW} \sim 10$, the relevant RGE running effects between these two
scales are negligibly small for the physics under consideration.},
%%%%%%%%%%%%%%%%%%%%
where $E_L^{}$ denotes the leptonic $SU(2)_L^{}$ doublets,
$H_1^{}$ and $H_2^{}$ are the Higgs fields, $l_R^{}$ denotes the
right-handed charged leptons, $H_2^c\equiv i \sigma_{}^2 H_2^\ast$
and $E_L^c\equiv i \sigma_{}^2 {\cal C} \overline{E_L^{}}_{}^T$
with ${\cal C}$ being the Dirac charge-conjugate matrix. After
spontaneous gauge symmetry breaking at the electroweak scale
$\Lambda_{\rm EW} \sim 10^2$ GeV, we arrive at the charged lepton
mass matrix $M_l = vY_l\cos\beta$ and the effective Majorana
neutrino mass matrix $M_\nu = v^2\kappa \sin^2\beta$, where $v
\approx 174$ GeV and $\tan\beta$ is the ratio of the vacuum
expectation values of $H_2$ and $H_1$ in the MSSM. The lepton
flavor mixing matrix $V$ arises from the mismatch between the
diagonalization of $Y_l$ (or $M_l$) and that of $\kappa$ (or
$M_\nu$). In the flavor basis where $Y_l$ is real and diagonal,
$V$ directly links the neutrino mass eigenstates $(\nu^{}_1,
\nu^{}_2, \nu^{}_3)$ to the neutrino flavor eigenstates
$(\nu^{}_e, \nu^{}_\mu, \nu^{}_\tau)$. The physical parameters at
$\Lambda_{\rm SS}$ and $\Lambda_{\rm EW}$ are related by the
renormalization group equations (RGEs) \cite{RGE}. It has been
shown that the RGE evolution between these two scales may have
significant effects on the mixing angle $\theta_{12}$ and the
CP-violating phases $\delta$, $\rho$ and $\sigma$
\cite{RGE1,RGE2,RGE3,RGE4,Mei}, in particular if the masses of
three light neutrinos are nearly degenerate. It has also been
noticed by Casas {\it et al} \cite{RGE2} and Antusch {\it et al}
\cite{RGE3} that a CP-violating phase can be radiatively generated
due to the RGE running from $\Lambda_{\rm SS}$ to $\Lambda_{\rm
EW}$.

The reason for the radiative generation of a CP-violating phase is
simply that three phases of $V$ are entangled with one another in
the RGEs. In other words, the running behavior of $\delta$ depends
on a non-linear function of $\delta$, $\rho$ and $\sigma$. It is
therefore possible to generate a non-zero value of $\delta$ at
$\Lambda_{\rm EW}$ even if $\delta =0$ holds at $\Lambda_{\rm
SS}$, provided the initial values of $\rho$ and $\sigma$ are not
vanishing. This observation opens a new and interesting window to
understand possible connection between the phenomena of CP
violation at low- and high-energy scales; e.g., the phase
parameter governing the strength of CP violation in a
long-baseline neutrino oscillation experiment could be radiatively
generated from those CP-violating phases which control the
leptogenesis of right-handed neutrinos \cite{LEP} at the seesaw
scale.

The main purpose of this paper is to analyze the radiative
generation of three CP-violating phases ($\delta, \rho, \sigma$)
via the one-loop RGE running effects from $\Lambda_{\rm SS}$ to
$\Lambda_{\rm EW}$. Our work is different from the one done in
Refs. \cite{RGE2} and \cite{RGE3} at least in the following
aspects:
\begin{itemize}
\item       The phase convention of $V$ taken by us in Eq. (1)
forbids the phase parameter $\delta$ to appear in the effective
mass of the neutrinoless double-beta decay $\langle
m\rangle_{ee}$. Hence it makes sense to refer to $\delta$ as the
Dirac phase. In contrast, the so-called ``Dirac" phase in the
``standard" parametrization of $V$ used by Casas {\it et al}
\cite{RGE2} and Antusch {\it et al} \cite{RGE3} enters the
expression of $\langle m\rangle_{ee}$ and is not purely of the
Dirac nature. Our RGEs for three CP-violating phases turn out to
be remarkably different from theirs.

\item       We focus on the possibilities to generate $\delta =
90^\circ$, $\rho = 90^\circ$ or $\sigma = 90^\circ$ at
$\Lambda_{\rm EW}$ from $\delta = 0^\circ$, $\rho =0^\circ$ or
$\sigma =0^\circ$ at $\Lambda_{\rm SS}$. While $\delta =90^\circ$
might imply ``Maximal" CP violation in some sense \cite{FX95},
$\rho =90^\circ$ or $\sigma = 90^\circ$ will lead to a kind of
large cancellation in $\langle m\rangle_{ee}$. The RGE running of
$\delta$ is of particular interest, because it means the radiative
generation of leptonic unitarity triangles in the complex plane;
i.e., three overlapped lines (sides) at $\Lambda_{\rm SS}$ can
evolve into a triangle at $\Lambda_{\rm EW}$.

\item       The one-loop RGEs for the Jarlskog parameter (denoted
by $\cal J$) \cite{J} and two off-diagonal asymmetries of $V$
(denoted as ${\cal A}^{}_{\rm L}$ and ${\cal A}^{}_{\rm R}$)
\cite{Xing02} are derived, and their running behaviors from
$\Lambda_{\rm SS}$ to $\Lambda_{\rm EW}$ are illustrated. These
rephasing-invariant quantities measure the strength of Dirac-type
CP violation and the geometrical structure of $V$, respectively.
For example, ${\cal A}^{}_{\rm L}=0$ (or ${\cal A}^{}_{\rm R}=0$)
would imply $V$ to be symmetric about its $V_{e1}$-$V_{\mu
2}$-$V_{\tau 3}$ (or $V_{e3}$-$V_{\mu 2}$-$V_{\tau 1}$) axis.
\end{itemize}
Because the radiative corrections to $V$ are expected to be
insignificant in the case that three light neutrinos have a strong
mass hierarchy and $\tan\beta$ takes small or mild values, we
shall mainly consider the quasi-degenerate neutrino mass spectrum
in our numerical calculations.

Section II is devoted to the one-loop RGEs for three neutrino
masses, three lepton flavor mixing angles, three CP-violating
phases and three rephasing-invariant quantities of $V$. A detailed
numerical analysis of radiative corrections to the Dirac and
Majorana phases is presented in section III, where the RGE
evolution of ${\cal J}$, ${\cal A}_{\rm L}$ and ${\cal A}_{\rm R}$
is also analyzed. We give a brief summary of our results with some
concluding remarks in Section IV.

\section{RGEs for physical parameters}

Below the seesaw scale, the effective neutrino coupling matrix
$\kappa$ obeys the following one-loop RGE in the MSSM:
\begin{equation}
16\pi^2 \frac{{\rm d}\kappa}{{\rm d}t} = \alpha \kappa + \left
(Y^{}_lY^\dagger_l \right ) \kappa + \kappa \left (Y^{}_l
Y^\dagger_l \right )^T \;
%       (3)
\end{equation}
with $t\equiv \ln (\mu/\Lambda_{\rm SS})$ and
\begin{equation}
\alpha \; = \; -\frac{6}{5}g^2_1 - 6g^2_2 + 6 \left (y^2_u + y^2_c
+ y^2_t \right ) \; ,
%       (4)
\end{equation}
where $g^{}_1$ and $g^{}_2$ denote the gauge couplings, and
$y^{}_f$ (for $f=u,c,t$) stand for the Yukawa couplings of up-type
quarks. In the flavor basis where $Y_l$ is real and diagonal, we
have $\kappa = V \overline{\kappa} V^T$ with $\overline{\kappa} =
{\rm Diag}\{\kappa^{}_1, \kappa^{}_2, \kappa^{}_3\}$. The neutrino
masses at $\Lambda_{\rm EW}$ read as $m^{}_i = v^2 \kappa^{}_i
\sin^2\beta$ (for $i=1,2,3$). It is then possible to derive the
RGEs for $(\kappa^{}_1, \kappa^{}_2, \kappa^{}_3)$, $(\theta_{12},
\theta_{23}, \theta_{13})$ and $(\delta, \rho, \sigma)$ from Eq.
(3), similar to the work done in Refs. \cite{RGE2,RGE3,RGE4,Mei}.

For simplicity, we neglect the small contributions of $y^{}_e$ and
$y^{}_\mu$ to Eq. (3) in our calculations. The RGEs of
$\kappa^{}_i$ (for $i=1,2,3$) turn out to be
\begin{eqnarray}
\frac{{\rm d} \kappa^{}_1}{{\rm d} t} & = &
\frac{\kappa^{}_1}{16\pi^2} \left [ \alpha + 2 y_{\tau}^2 \left
(s_{12}^2 s_{23}^2 - 2c_{\delta }^{} c_{12}^{} c_{23}^{} s_{12}^{}
s_{23}^{} s_{13}^{} + c_{12}^2 c_{23}^2 s_{13}^2 \right ) \right ]
\; ,
\nonumber \\
\frac{{\rm d} \kappa^{}_2}{{\rm d} t} & = &
\frac{\kappa^{}_2}{16\pi^2} \left [ \alpha + 2 y_{\tau}^2 \left (
c_{12}^2 s_{23}^2 + 2c_{\delta }^{} c_{12}^{} c_{23}^{} s_{12}^{}
s_{23}^{} s_{13}^{} + c_{23}^2 s_{12}^2 s_{13}^2 \right ) \right ]
\; ,
\nonumber \\
\frac{{\rm d} \kappa^{}_3}{{\rm d} t} & = &
\frac{\kappa^{}_3}{16\pi^2} \left [ \alpha + 2 y_{\tau }^2
c_{23}^2 c_{13}^2 \right ] \; ,
%       (5)
\end{eqnarray}
where $c^{}_\delta \equiv \cos\delta$. The RGEs of $\theta_{ij}$
(for $ij =12, 23, 13$) are found to be
\begin{eqnarray}
\frac{{\rm d} \theta^{}_{12}}{{\rm d} t} & = &
\frac{y^2_\tau}{16\pi^2} \left \{\frac{c_{(\rho -\sigma
)}^{}}{\zeta_{12}^{}} \left [c_{(\rho -\sigma )}^{} c_{12}^{}
s_{12}^{} \left (s_{23}^2 - c_{23}^2 s_{13}^2 \right ) - \left
(c_{(\delta +\rho -\sigma)}^{} c_{12}^2 - c_{(\delta -\rho +\sigma
)}^{} s_{12}^2 \right ) c_{23}^{} s_{23}^{} s_{13}^{} \right ]
\right .
\nonumber \\
&& + \zeta_{12}^{} s_{(\rho -\sigma)}^{} \left [ s_{(\rho -\sigma
)}^{} c_{12}^{} s_{12}^{} \left (s_{23}^2 - c_{23}^2 s_{13}^2
\right ) - \left (s_{(\delta +\rho -\sigma )}^{} c_{12}^2 +
s_{(\delta -\rho +\sigma) }^{} s_{12}^2 \right ) c_{23}^{}
s_{23}^{} s_{13}^{} \right ]
\nonumber \\
&& - \left [\frac{c_{\rho}^{}}{\zeta_{13}^{}} \left (c_{(\delta
-\rho)}^{} s_{12}^{} s_{23}^{} - c_{\rho }^{} c_{12}^{} c_{23}^{}
s_{13}^{} \right ) - \zeta_{13}^{} s_{\rho}^{} \left (s_{(\delta
-\rho)}^{} s_{12}^{} s_{23}^{} + s_{\rho }^{} c_{12}^{} c_{23}^{}
s_{13}^{} \right ) \right ] c_{23}^{} s_{12}^{} s_{13}^{}
\nonumber \\
&& \left . - \left [\frac{c_{\sigma}^{}}{\zeta _{23}^{}} \left
(c_{(\delta -\sigma) }^{} c_{12}^{} s_{23}^{} + c_{\sigma}^{}
c_{23}^{} s_{12}^{} s_{13}^{} \right ) - \zeta _{23}^{}
s_{\sigma}^{} \left (s_{(\delta -\sigma) }^{} c_{12}^{} s_{23}^{}
- s_{\sigma }^{} c_{23}^{} s_{12}^{} s_{13}^{} \right ) \right ]
c_{12}^{} c_{23}^{} s_{13}^{} \right \} \; ,
\nonumber \\
\nonumber \\
\frac{{\rm d} \theta^{}_{23}}{{\rm d} t} & = &
\frac{y^2_\tau c^{}_{23}}{16\pi^2} \left \{ \left
[\frac{c_{(\delta -\rho )}^{}}{\zeta_{13}^{}} \left (c_{(\delta
-\rho )}^{} s_{12}^{} s_{23}^{} - c_{\rho }^{} c_{12}^{} c_{23}^{}
s_{13}^{} \right ) + \zeta_{13}^{} s_{(\delta -\rho)}^{} \left
(s_{(\delta -\rho) }^{} s_{12}^{} s_{23}^{} + s_{\rho}^{}
c_{12}^{} c_{23}^{} s_{13}^{} \right ) \right ] s_{12}^{} \right .
\nonumber \\
&& \left. + \left [\frac{c_{(\delta -\sigma )}^{}}{\zeta_{23}^{}}
\left (c_{(\delta -\sigma) }^{} c_{12}^{} s_{23}^{} +
c_{\sigma}^{} c_{23}^{} s_{12}^{} s_{13}^{} \right ) + \zeta
_{23}^{} s_{(\delta -\sigma)}^{} \left (s_{(\delta -\sigma )}^{}
c_{12}^{} s_{23}^{} - s_{\sigma }^{} c_{23}^{} s_{12}^{} s_{13}^{}
\right ) \right ] c_{12}^{} \right \} \; ,
\nonumber \\
\nonumber \\
\frac{{\rm d} \theta^{}_{13}}{{\rm d} t} & = &
\frac{y^2_\tau c^{}_{23}}{16\pi^2} \left \{- \left [
\frac{c_{\rho}^{}}{\zeta_{13}^{}} \left (c_{(\delta -\rho )}^{}
s_{12}^{} s_{23}^{} - c_{\rho}^{} c_{12}^{} c_{23}^{} s_{13}^{}
\right ) - \zeta_{13}^{} s_{\rho}^{} \left (s_{(\delta -\rho) }^{}
s_{12}^{} s_{23}^{} + s_{\rho }^{} c_{12}^{} c_{23}^{} s_{13}^{}
\right ) \right ] c_{12}^{} c_{13}^{} \right.
\nonumber \\
&& \left . + \left [\frac{c_{\sigma}^{}}{\zeta _{23}^{}}
\left(c_{(\delta -\sigma) }^{} c_{12}^{} s_{23}^{} + c_{\sigma}^{}
c_{23}^{} s_{12}^{} s_{13}^{} \right ) - \zeta _{23}^{}
s_{\sigma}^{} \left(s_{(\delta -\sigma) }^{} c_{12}^{} s_{23}^{} -
s_{\sigma }^{} c_{23}^{} s_{12}^{} s_{13}^{} \right ) \right ]
c_{13}^{} s_{12}^{} \right \} \; ,
%       (6)
\end{eqnarray}
in which $\zeta^{}_{ij} \equiv (\kappa^{}_i -
\kappa^{}_j)/(\kappa^{}_i + \kappa^{}_j)$ (for $ij = 12, 23, 13$),
$c^{}_A \equiv \cos A$ and $s^{}_A \equiv \sin A$ (for $A =
\delta, ~\rho, ~\sigma, ~\delta - \rho, ~\delta - \sigma, ~\rho -
\sigma, ~\delta + \rho - \sigma, ~\delta - \rho + \sigma$). In
addition, the RGEs of $\delta$, $\rho$ and $\sigma$ are obtained
as follows:
\begin{eqnarray}
\frac{{\rm d} \delta}{{\rm d} t} & = & \frac{y^2_\tau}{16\pi^2}
\left \{\frac{s_{(\rho -\sigma)}^{}}{\zeta_{12}^{}} \left [
c_{(\rho -\sigma)}^{} \left (s_{23}^2 - c_{23}^2 s_{13}^2 \right )
- \left (c_{(\delta +\rho -\sigma)}^{} c_{12}^2 - c_{(\delta -\rho
+ \sigma)}^{} s_{12}^2 \right ) \frac{c_{23}^{} s_{23}^{}
s_{13}^{}}{c_{12}^{} s_{12}^{}} \right ] \right .
\nonumber \\
&& - \zeta_{12}^{} c_{(\rho -\sigma)}^{} \left [s_{(\rho -\sigma
)}^{} \left (s_{23}^2 - c_{23}^2 s_{13}^2 \right ) - \left
(s_{(\delta +\rho -\sigma)}^{} c_{12}^2 + s_{(\delta -\rho +\sigma
)}^{} s_{12}^2 \right ) \frac{c_{23}^{} s_{23}^{}
s_{13}^{}}{c_{12}^{} s_{12}^{}} \right ]
\nonumber \\
&& + \frac{1}{\zeta_{13}} \left (c_{(\delta -\rho )} s_{12}^{}
s_{23}^{} - c_{\rho}^{} c_{12}^{} c_{23}^{} s_{13}^{} \right )
\left [ \frac{s_{\rho}^{} c_{23}^{}}{c_{12}^{} s_{13}^{}} \left
(c_{12}^2 - s_{12}^2 s_{13}^2 \right ) - s_{(\delta -\rho )}^{}
\frac{s_{12}^{}}{s_{23}^{}} \left (c_{23}^2 - s_{23}^2 \right )
\right ]
\nonumber \\
&& + \zeta_{13}^{} \left (s_{(\delta -\rho )}^{} s_{12}^{}
s_{23}^{} + s_{\rho} c_{12}^{} c_{23}^{} s_{13}^{} \right ) \left
[ \frac{c_{\rho}^{} c_{23}^{}}{c_{12}^{} s_{13}^{}} \left
(c_{12}^2 - s_{12}^2 s_{13}^2 \right ) + c_{(\delta -\rho )}^{}
\frac{s_{12}^{}}{s_{23}^{}}
\left (c_{23}^2 - s_{23}^2 \right ) \right ]
\nonumber \\
&& - \frac{1}{\zeta_{23}} \left (c_{(\delta -\sigma )} c_{12}^{}
s_{23}^{} + c_{\sigma}^{} c_{23}^{} s_{12}^{} s_{13}^{} \right )
\left [\frac{s_{\sigma}^{} c_{23}^{}}{s_{12}^{} s_{13}^{}}
\left(s_{12}^2 - c_{12}^2 s_{13}^2 \right ) + s_{(\delta -\sigma
)}^{} \frac{c_{12}^{}}{s_{23}^{}} \left (c_{23}^2 - s_{23}^2
\right ) \right ]
\nonumber \\
&& \left . - \zeta_{23}^{} \left (s_{(\delta -\sigma )}^{}
c_{12}^{} s_{23}^{} - s_{\sigma} c_{23}^{} s_{12}^{} s_{13}^{}
\right ) \left [\frac{c_{\sigma}^{} c_{23}^{}}{s_{12}^{}
s_{13}^{}} \left(s_{12}^2 - c_{12}^2 s_{13}^2 \right ) -
c_{(\delta -\sigma)}^{} \frac{c_{12}^{}}{s_{23}^{}}
\left (c_{23}^2 - s_{23}^2 \right ) \right ] \right \} \; ,
\nonumber \\
\nonumber \\
\frac{{\rm d} \rho}{{\rm d} t} & = & \frac{y^2_\tau}{16\pi^2}
\left \{\frac{s_{(\rho -\sigma)}^{}}{\zeta_{12}^{}} \left [
c_{(\rho -\sigma )}^{} c_{12}^{} s_{12}^{} \left(s_{23}^2 -
c_{23}^2s_{13}^2 \right ) - \left(c_{(\delta +\rho -\sigma )}^{}
c_{12}^2 - c_{(\delta -\rho +\sigma)}^{} s_{12}^2 \right )
c_{23}^{} s_{23}^{} s_{13}^{} \right ] \frac{s_{12}^{}}{c_{12}^{}}
\right .
\nonumber \\
&& - \zeta_{12}^{} c_{(\rho -\sigma)}^{} \left [s_{(\rho -\sigma
)}^{} c_{12}^{} s_{12}^{} \left (s_{23}^2 - c_{23}^2 s_{13}^2
\right ) - \left (s_{(\delta +\rho -\sigma )}^{} c_{12}^2 +
s_{(\delta -\rho +\sigma)}^{} s_{12}^2 \right ) c_{23}^{}
s_{23}^{} s_{13}^{} \right ] \frac{s_{12}^{}}{c_{12}^{}}
\nonumber \\
&& + \left [\frac{s_{\rho}^{}}{\zeta_{13}^{}} \left (c_{(\delta
-\rho)}^{} s_{12}^{} s_{23}^{} - c_{\rho }^{} c_{12}^{} c_{23}^{}
s_{13}^{} \right ) + \zeta_{13}^{} c_{\rho}^{} \left (s_{(\delta
-\rho)}^{} s_{12}^{} s_{23}^{} + s_{\rho } c_{12}^{} c_{23}^{}
s_{13}^{} \right ) \right ]
\frac{c_{23}^{} \left (c_{12}^2 c_{13}^2 - s_{13}^2 \right )}{c_{12}^{}s_{13}^{}}
\nonumber \\
&& \left . - \left [\frac{s_{\sigma}^{}}{\zeta _{23}^{}} \left
(c_{(\delta -\sigma) }^{} c_{12}^{} s_{23}^{} + c_{\sigma}^{}
c_{23}^{} s_{12}^{} s_{13}^{} \right ) + \zeta _{23}^{}
c_{\sigma}^{} \left (s_{(\delta -\sigma) }^{} c_{12}^{} s_{23}^{}
- s_{\sigma} c_{23}^{} s_{12}^{} s_{13}^{} \right ) \right ]
\frac{c_{23}^{} c_{13}^2 s_{12}^{}}{s_{13}^{}} \right \} \; ,
\nonumber \\
\nonumber \\
\frac{{\rm d} \sigma}{{\rm d} t} & = & \frac{y^2_\tau}{16\pi^2}
\left \{\frac{s_{(\rho -\sigma)}^{}}{\zeta_{12}^{}} \left [
c_{(\rho -\sigma )}^{} c_{12}^{} s_{12}^{} \left (s_{23}^2 -
c_{23}^2 s_{13}^2 \right ) - \left (c_{(\delta +\rho -\sigma )}^{}
c_{12}^2 - c_{(\delta -\rho +\sigma)}^{} s_{12}^2 \right )
c_{23}^{} s_{23}^{} s_{13}^{} \right ] \frac{c_{12}^{}}{s_{12}^{}}
\right .
\nonumber \\
&& - \zeta_{12}^{} c_{(\rho -\sigma)}^{} \left [s_{(\rho -\sigma
)}^{} c_{12}^{} s_{12}^{} \left (s_{23}^2 - c_{23}^2 s_{13}^2
\right ) - \left (s_{(\delta +\rho -\sigma )}^{} c_{12}^2 +
s_{(\delta -\rho +\sigma)}^{} s_{12}^2 \right ) c_{23}^{}
s_{23}^{} s_{13}^{} \right ] \frac{c_{12}^{}}{s_{12}^{}}
\nonumber \\
&& - \left [\frac{s_{\sigma}^{}}{\zeta _{23}^{}} \left (c_{(\delta
-\sigma) }^{} c_{12}^{} s_{23}^{} + c_{\sigma}^{} c_{23}^{}
s_{12}^{} s_{13}^{} \right ) + \zeta _{23}^{} c_{\sigma}^{} \left
(s_{(\delta -\sigma) }^{} c_{12}^{} s_{23}^{} - s_{\sigma}
c_{23}^{} s_{12}^{} s_{13}^{} \right ) \right ] \frac{c_{23}^{}
\left (c_{13}^2 s_{12}^2 - s_{13}^2 \right ) }{s_{12}^{}
s_{13}^{}}
\nonumber \\
&& \left . + \left [\frac{s_{\rho}^{}}{\zeta_{13}^{}} \left
(c_{(\delta -\rho)}^{} s_{12}^{} s_{23}^{} - c_{\rho }^{}
c_{12}^{} c_{23}^{} s_{13}^{} \right ) + \zeta_{13}^{} c_{\rho}^{}
\left (s_{(\delta -\rho)}^{} s_{12}^{} s_{23}^{} + s_{\rho }
c_{12}^{} c_{23}^{} s_{13}^{} \right ) \right ] \frac{c_{12}^{}
c_{23}^{} c_{13}^2}{s_{13}^{}} \right \} \; .
%       (7)
\end{eqnarray}
It is worth mentioning that the analytical results in Eqs. (6) and
(7) can also be achieved from Eqs. (14)--(19) of Ref. \cite{Mei}
by setting $C^l_\kappa =1$ and $y^{}_\nu =0$ over there. One can
see that the running effects of three flavor mixing angles and
three CP-violating phases are all proportional to the factor
$y^2_\tau/(16\pi^2) = m^2_\tau(1+\tan^2\beta)/(16\pi^2v^2) \approx
6.6\times 10^{-7} (1+\tan^2\beta)$ in the MSSM. Hence an
appreciable value of $\tan\beta$ is required, in order to get
appreciable radiative corrections to relevant physical parameters.

Next let us consider three rephasing-invariant quantities of $V$.
The first one is the Jarlskog parameter $\cal J$ \cite{J}, defined
through
\begin{equation}
{\rm Im} \left (V_{\alpha i}V_{\beta j} V^*_{\alpha j}V^*_{\beta
i} \right ) = {\cal J} \sum_{\gamma,k} \left (\epsilon^{~}_{\alpha
\beta \gamma} \epsilon^{~}_{ijk} \right ) \; ,
%       (8)
\end{equation}
where the Greek and Latin subscripts run over $(e,\mu,\tau)$ and
$(1,2,3)$, respectively. Taking account of the parametrization of
$V$ in Eq. (1), we have ${\cal J} = s^{}_{12} c^{}_{12} s^{}_{23}
c^{}_{23} s^{}_{13} c^2_{13} s^{}_\delta$. The off-diagonal
asymmetries of $V$ \cite{Xing02},
\begin{eqnarray}
{\cal A}^{}_{\rm L} & \equiv & |V_{e2}|^2 - |V_{\mu 1}|^2 =
|V_{\mu 3}|^2 - |V_{\tau 2}|^2 = |V_{\tau 1}|^2 - |V_{e 3}|^2 \; ,
\nonumber \\
{\cal A}^{}_{\rm R} & \equiv & |V_{e2}|^2 - |V_{\mu 3}|^2 =
|V_{\mu 1}|^2 - |V_{\tau 2}|^2 = |V_{\tau 3}|^2 - |V_{e 1}|^2 \; ,
%       (9)
\end{eqnarray}
are also rephasing-invariant. We obtain ${\cal A}^{}_{\rm L} =
s^2_{12} \left (c^2_{13} - c^2_{23} \right ) - c^2_{12} s^2_{23}
s^2_{13} - 2 s^{}_{12} c^{}_{12} s^{}_{23} c^{}_{23} s^{}_{13}
c^{}_\delta$ and ${\cal A}^{}_{\rm R} = c^2_{13} \left (s^2_{12} -
s^2_{23} \right )$. The RGEs of $\cal J$, ${\cal A}^{}_{\rm L}$
and ${\cal A}^{}_{\rm R}$ can then be derived from Eqs. (6) and
(7). The result for $\cal J$ is
\begin{eqnarray}
\frac{{\rm d} {\cal J}}{{\rm d} t} & = & \frac{y^2_\tau}{16\pi^2}
\left \{- \frac{c_{23}^{} c_{13}^2 s_{23}^{}
s_{13}^{}}{\zeta_{12}^{}} \left [c_\delta^{} s_{(\rho-\sigma)}^{}
+ c_{(\rho-\sigma)}^{} s_\delta^{} \left (c_{12}^2 - s_{12}^2
\right ) \right ] \left [c_{(\delta+\rho-\sigma)}^{} c_{12}^2
c_{23}^{} s_{23}^{} s_{13}^{} \right . \right .
\nonumber \\
&& \left. - c_{(\delta-\rho+\sigma)}^{} c_{23}^{} s_{12}^2
s_{23}^{} s_{13}^{} - c_{(\rho-\sigma)}^{} c_{12}^{} s_{12}^{}
\left (s_{23}^2 - c_{23}^2s_{13}^2 \right ) \right ]
\nonumber \\
&& + \zeta_{12}^{} c_{23}^{} c_{13}^2 s_{23}^{} s_{13}^{} \left
[c_\delta^{} c_{(\rho-\sigma)}^{} - s_\delta^{}
s_{(\rho-\sigma)}^{} \left (c_{12}^2 - s_{12}^2 \right ) \right ]
\left [ s_{(\delta+\rho-\sigma)}^{} c_{12}^2 c_{23}^{} s_{23}^{}
s_{13}^{} \right .
\nonumber \\
&& \left . + s_{(\delta-\rho+\sigma)}^{} c_{23}^{} s_{12}^2
s_{23}^{} s_{13}^{} - s_{(\rho-\sigma)}^{} c_{12}^{} s_{12}^{}
\left (s_{23}^2 - c_{23}^2 s_{13}^2 \right ) \right ]
\nonumber \\
&& + \frac{c_{23}^{} c_{13}^2 s_{12}^{}}{\zeta_{13}^{}} \left
(c_{(\delta-\rho)}^{} s_{12}^{} s_{23}^{} - c_\rho^{} c_{12}^{}
c_{23}^{} s_{13}^{} \right ) \left [s_\rho^{} c_{12}^{} s_{12}^{}
s_{13}^{} \left (c_{23}^2 - s_{23}^2 \right ) \right .
\nonumber\\
&& \left . + s_{(\delta-\rho)}^{} c_{23}^{} s_{12}^2 s_{23}^{}
s_{13}^2 + \left [c_\delta^{} s_\rho^{} - c_\rho^{} s_\delta^{}
\left (c_{13}^2 - s_{13}^2 \right ) \right ] c_{12}^2 c_{23}^{}
s_{23}^{} \right ]
\nonumber \\
&& + \zeta_{13}^{} c_{23}^{} c_{13}^2 s_{12}^{} \left
(s_{(\delta-\rho)}^{} s_{12}^{} s_{23}^{} + s_\rho^{} c_{12}^{}
c_{23}^{} s_{13}^{} \right ) \left [c_\rho^{} c_{12}^{} s_{12}^{}
s_{13}^{} \left (c_{23}^2 - s_{23}^2 \right ) \right .
\nonumber \\
&& \left . - c_{(\delta-\rho)}^{} c_{23}^{} s_{12}^2 s_{23}^{}
s_{13}^2 + \left [c_\delta^{} c_\rho^{} + s_\delta^{} s_\rho^{}
\left (c_{13}^2 - s_{13}^2 \right ) \right ] c_{12}^2 c_{23}^{}
s_{23}^{} \right ]
\nonumber \\
&& - \frac{c_{12}^{} c_{23}^{} c_{13}^2}{\zeta_{23}^{}} \left
(c_{(\delta-\sigma)}^{} c_{12}^{} s_{23}^{} + c_\sigma^{}
c_{23}^{} s_{12}^{} s_{13}^{} \right ) \left [ \left (c_\delta^{}
s_\sigma^{} - c_\sigma^{} s_\delta^{} c_{13}^2 \right ) c_{23}^{}
s_{12}^2 s_{23}^{} \right .
\nonumber \\
&& \left . - s_\sigma^{} c_{12}^{} s_{12}^{} s_{13}^{} \left
(c_{23}^2 - s_{23}^2 \right ) + \left (s_{(\delta-\sigma)}^{}
c_{12}^2 + c_\sigma^{} s_\delta^{} s_{12}^2 \right ) c_{23}^{}
s_{23}^{} s_{13}^2 \right ]
\nonumber \\
&& + \zeta_{23}^{} c_{12}^{} c_{23}^{} c_{13}^2 \left
(s_{(\delta-\sigma)}^{} c_{12}^{} s_{23}^{} - s_\sigma^{}
c_{23}^{} s_{12}^{} s_{13}^{} \right ) \left [- \left (c_\delta^{}
c_\sigma^{} + s_\delta^{} s_\sigma^{} c_{13}^2 \right ) c_{23}^{}
s_{12}^2 s_{23}^{} \right .
\nonumber \\
&& + c_\sigma^{} c_{12}^{} s_{12}^{} s_{13}^{} \left (c_{23}^2 -
s_{23}^2 \right ) + \left . \left . \left (c_{(\delta-\sigma)}^{}
c_{12}^2 + s_\delta^{} s_\sigma^{} s_{12}^2 \right ) c_{23}^{}
s_{23}^{} s_{13}^2 \right ] \right \} \; .
%       (10)
\end{eqnarray}
Different from ${\rm d}\delta/{\rm d} t$, ${\rm d} {\cal J}/{\rm
d} t$ does not suffer from any divergence in the $\theta_{13}
\rightarrow 0$ limit. This feature proves that ${\cal J}$ itself
is a well-defined rephasing-invariant quantity, while $\delta$ is
parametrization-dependent and cannot be well defined when
$\theta_{13}$ vanishes. The RGEs of ${\cal A}^{}_{\rm L}$ and
${\cal A}^{}_{\rm R}$ read as follows:
\begin{eqnarray}
\frac{{\rm d}{\cal A}_{\rm L}^{}}{{\rm d} t} & = &
\frac{y_\tau^2}{16\pi^2} \left \{ \frac{2}{\zeta_{12}^{}} \left
[c_{(\delta+\rho-\sigma)}^{} c_{12}^2 c_{23}^{} s_{23}^{}
s_{13}^{} - c_{(\delta-\rho+\sigma)}^{} c_{23}^{} s_{12}^2
s_{23}^{} s_{13}^{} - c_{(\rho-\sigma)}^{} c_{12}^{} s_{12}^{}
\left (s_{23}^2 - c_{23}^2 s_{13}^2 \right ) \right ] \right.
\nonumber \\
&& \left [-s_\delta^{} s_{(\rho-\sigma)}^{} c_{23}^{} s_{23}^{}
s_{13}^{} + c_\delta^{} c_{(\rho-\sigma)}^{} c_{23}^{} s_{23}^{}
s_{13}^{} \left (c_{12}^2 - s_{12}^2 \right ) -
c_{(\rho-\sigma)}^{} c_{12}^{} s_{12}^{} \left (s_{23}^2 -
c_{23}^2 s_{13}^2 \right ) \right ]
\nonumber \\
&& + 2\zeta_{12}^{} \left [s_{(\delta+\rho-\sigma)}^{} c_{12}^2
c_{23}^{} s_{23}^{} s_{13}^{} + s_{(\delta-\rho+\sigma)}^{}
c_{23}^{} s_{12}^2 s_{23}^{} s_{13}^{} - s_{(\rho-\sigma)}^{}
c_{12}^{} s_{12}^{} \left (s_{23}^2 - c_{23}^2 s_{13}^2 \right )
\right ]
\nonumber \\
&& \left [c_{(\rho-\sigma)}^{} s_\delta^{} c_{23}^{} s_{23}^{}
s_{13}^{} + c_\delta^{} s_{(\rho-\sigma)}^{} c_{23}^{} \left
(c_{12}^2 - s_{12}^2 \right ) s_{23}^{} s_{13}^{} -
s_{(\rho-\sigma)}^{} c_{12}^{} s_{12}^{} \left (s_{23}^2 -
c_{23}^2 s_{13}^2 \right ) \right ]
\nonumber \\
&& + \frac{2c_{23}^{} c_{13}^2 s_{23}^{}}{\zeta_{13}^{}} \left
(c_{(\delta-\rho)}^{} s_{12}^{} s_{23}^{} - c_\rho^{} c_{12}^{}
c_{23}^{} s_{13}^{} \right ) \left (c_{(\delta-\rho)}^{} c_{23}^{}
s_{12}^{} + c_\rho^{} c_{12}^{} s_{23}^{} s_{13}^{} \right )
\nonumber \\
&& + 2\zeta_{13}^{} c_{23}^{} c_{13}^2 s_{23}^{} \left
(s_{(\delta-\rho)}^{} s_{12}^{} s_{23}^{} + s_\rho^{} c_{12}^{}
c_{23}^{} s_{13}^{} \right ) \left (s_{(\delta-\rho)}^{} c_{23}^{}
s_{12}^{} - s_\rho^{} c_{12}^{} s_{23}^{} s_{13}^{} \right )
\nonumber \\
&& - \frac{2c_\sigma^{} c_{23}^{} c_{13}^2 s_{12}^{}
s_{13}^{}}{\zeta_{23}^{}} \left (c_{(\delta-\sigma)}^{} c_{12}^{}
s_{23}^{} + c_\sigma^{} c_{23}^{} s_{12}^{} s_{13}^{} \right )
\nonumber \\
&& \left . + 2\zeta_{23}^{} s_\sigma^{} c_{23}^{} c_{13}^2
s_{12}^{} s_{13}^{} \left (s_{(\delta-\sigma)}^{} c_{12}^{}
s_{23}^{} - s_\sigma^{} c_{23}^{} s_{12}^{} s_{13}^{} \right )
\right \} \; ,
\nonumber \\
\nonumber \\
\frac{{\rm d}{\cal A}_{\rm R}^{}}{{\rm d} t} & = & \frac{y_\tau^2
c^2_{13}}{16\pi^2} \left \{ \frac{2c_{(\rho-\sigma)}^{} c_{12}^{}
s_{12}^{}}{\zeta_{12}^{}} \left [-c_{(\delta+\rho-\sigma)}^{}
c_{12}^2 c_{23}^{} s_{23}^{} s_{13}^{} +
c_{(\delta-\rho+\sigma)}^{} c_{23}^{} s_{12}^2 s_{23}^{} s_{13}^{}
\right . \right .
\nonumber \\
&& \left . + c_{(\rho-\sigma)}^{} c_{12}^{} s_{12}^{} \left
(s_{23}^2 - c_{23}^2 s_{13}^2 \right ) \right ] - 2\zeta_{12}^{}
s_{(\rho-\sigma)}^{} c_{12}^{} s_{12}^{} \left
[s_{(\delta+\rho-\sigma)}^{} c_{12}^2 c_{23}^{} s_{23}^{}
s_{13}^{} \right .
\nonumber \\
&& \left . + s_{(\delta-\rho+\sigma)}^{} c_{23}^{} s_{12}^2
s_{23}^{} s_{13}^{} - s_{(\rho-\sigma)}^{} c_{12}^{} s_{12}^{}
\left (s_{23}^2 - c_{23}^2 s_{13}^2 \right ) \right ]
\nonumber \\
&& + \frac{2c_{23}^{} s_{23}^{}}{\zeta_{13}^{}} \left
(-c_{(\delta-\rho)}^{} s_{12}^{} s_{23}^{} + c_\rho^{} c_{12}^{}
c_{23}^{} s_{13}^{} \right ) \left (c_{(\delta-\rho)}^{} c_{23}^{}
s_{12}^{} + c_\rho^{} c_{12}^{} s_{23}^{} s_{13}^{} \right )
\nonumber \\
&& + 2\zeta_{13}^{} c_{23}^{} s_{23}^{} \left
(s_{(\delta-\rho)}^{} s_{12}^{} s_{23}^{} + s_\rho^{} c_{12}^{}
c_{23}^{} s_{13}^{} \right ) \left (-s_{(\delta-\rho)}^{}
c_{23}^{} s_{12}^{} + s_\rho^{} c_{12}^{} s_{23}^{} s_{13}^{}
\right )
\nonumber \\
&& - \frac{2c_{23}^2}{\zeta_{23}^{}} \left (c_{(\delta-\sigma)}^{}
c_{12}^{} s_{23}^{} + c_\sigma^{} c_{23}^{} s_{12}^{} s_{13}^{}
\right )^2 - \left . 2\zeta_{23}^{} c_{23}^2 \left
(s_{(\delta-\sigma)}^{} c_{12}^{} s_{23}^{} - s_\sigma^{}
c_{23}^{} s_{12}^{} s_{13}^{} \right )^2 \right \} \; .
%       (11)
\end{eqnarray}
Eqs. (10) and (11) clearly show that the RGE evolution of $\cal
J$, ${\cal A}^{}_{\rm L}$ and ${\cal A}^{}_{\rm R}$ depends on the
Majorana phases $\rho$ and $\sigma$. Hence the radiative
corrections to these rephasing-invariant parameters would in
general be quite different, if neutrinos were Dirac particles
instead of Majorana particles
%%%%%%%%%%%%%%%%%%%%%
\footnote{See, e.g., Ref. \cite{Lindner} for a recent analysis of
the RGE evolution of Dirac neutrino masses and lepton flavor
mixing parameters.}.
%%%%%%%%%%%%%%%%%%%%%

\section{Numerical Examples and Discussion}

We proceed to illustrate the radiative generation of three
CP-violating phases by taking a few typical numerical examples.
The eigenvalues of $Y_l$ at $\Lambda_{\rm SS}$ are chosen in such
a way that they can correctly run to their low-energy values
\cite{PDG}
%%%%%%%%%%%%%%%%%%%%%%%%%%%
\footnote{A similar treatment is required for the gauge couplings,
the quark Yukawa coupling eigenvalues and the quark flavor mixing
parameters in the full set of RGEs
\cite{RGE1,RGE2,RGE3,RGE4,Mei,FX00}.}.
%%%%%%%%%%%%%%%%%%%%%%%%%%%%
We assume the masses of three light neutrinos to be nearly
degenerate and $m^{}_1 \sim 0.2$ eV, so as to make the RGE running
effects of relevant physical quantities significant enough. The
initial values of $\kappa^{}_1$, $\kappa^{}_2$ and $\kappa^{}_3$
can be adjusted via
\begin{eqnarray}
\kappa^{}_1 & = & \frac{m^{}_1}{v^2\sin^2\beta} \; ,
\nonumber \\
\kappa^{}_2 & = & \frac{\sqrt{m^2_1 + \Delta
m^2_{21}}}{v^2\sin^2\beta} \; ,
\nonumber \\
\kappa^{}_3 & = & \frac{\sqrt{m^2_1 + \Delta
m^2_{31}}}{v^2\sin^2\beta} \; ,
%       (12)
\end{eqnarray}
together with a typical input $\tan\beta = 10$, such that the
resultant neutrino mass-squared differences $\Delta m^2_{21}
\equiv m^2_2 - m^2_1$ and $\Delta m^2_{31} \equiv m^2_3 - m^2_1$
at $\Lambda_{\rm EW}$ are consistent with the solar and
atmospheric neutrino oscillation data \cite{SNO,SK,KM,CHOOZ,K2K}.
We follow a similar strategy to choose the initial values of three
mixing angles $\theta_{12}$, $\theta_{23}$ and $\theta_{13}$, in
order to reproduce their low-energy values determined or
constrained from a global analysis of current experimental data
\cite{Fit}. In view of the upper bound $\theta_{13} < 10^\circ$,
we shall typically take $\theta_{13} = 1^\circ$, $3^\circ$ and
$5^\circ$ in our numerical calculations. We allow one of three
CP-violating phases ($\delta$, $\rho$, $\sigma$) to vanish at
$\Lambda_{\rm SS}$ and examine whether it can run to $90^\circ$ at
$\Lambda_{\rm EW}$ by choosing the initial values of the other two
phase parameters properly.

\subsection{Radiative Generation of $\delta=90^\circ$}

First of all, let us look at the one-loop RGE evolution of
$\delta$ in the $m^{}_1 < m^{}_2 < m^{}_3$ case (i.e., $\Delta
m^2_{31} > 0$). The input and output values of relevant physical
parameters are listed in Table I. One can see that $\delta =
90^\circ$ at $\Lambda_{\rm EW}$ can be radiatively generated from
$\delta = 0^\circ$ at $\Lambda_{\rm SS}$, if $\theta_{13} =
1^\circ$, $\rho = 4.0^\circ$ and $\sigma = -57.5^\circ$ are input.
Changing the initial value of $\theta_{13}$ to $3^\circ$ or
$5^\circ$ but fixing the input values of the other quantities, we
find that only $\delta = 41.8^\circ$ or $\delta = 35.8^\circ$ can
be obtained at $\Lambda_{\rm EW}$. While the results of $m^{}_1$,
$\Delta m^2_{31}$, $\theta_{12}$ and $\theta_{23}$ at
$\Lambda_{\rm EW}$ are rather stable against the change of
$\theta_{13}$ from $1^\circ$ to $5^\circ$ at $\Lambda_{\rm SS}$,
the result of $\Delta m^2_{21}$ becomes smaller and less favored.

The running behaviors of $\delta$, $\rho$ and $\sigma$ are
explicitly shown in FIG. 1, in which those of ${\cal J}$, ${\cal
A}^{}_{\rm L}$ and ${\cal A}^{}_{\rm R}$ are also illustrated.
Some comments are in order.
\begin{itemize}
\item       The one-loop RGE running behaviors of three
CP-violating phases are quite similar in the chosen parameter
space. It is easy to understand this feature from Eq. (7): given
the conditions that $\kappa^{}_1 \approx \kappa^{}_2 \approx
\kappa^{}_3$ holds and $\theta_{13}$ is small, Eq. (7) can be
simplified to
\begin{eqnarray}
\frac{{\rm d} \delta}{{\rm d} t} & \approx &
\frac{y^2_\tau}{16\pi^2} \left [\frac{c^{}_{(\rho -\sigma)}
s^{}_{(\rho -\sigma)}}{\zeta^{}_{12}} s^2_{23} + \left (
\frac{c^{}_{(\delta -\rho)} s^{}_\rho}{\zeta^{}_{13}} -
\frac{c^{}_{(\delta -\sigma)} s^{}_\sigma}{\zeta^{}_{23}} \right )
\frac{c^{}_{12} s^{}_{12} c^{}_{23} s^{}_{23}}{s^{}_{13}} \right ]
\; ,
\nonumber \\
\frac{{\rm d} \rho}{{\rm d} t} & \approx &
\frac{y^2_\tau}{16\pi^2} \left [\frac{c^{}_{(\rho -\sigma)}
s^{}_{(\rho -\sigma)}}{\zeta^{}_{12}} s^2_{12} s^2_{23} + \left (
\frac{c^{}_{(\delta -\rho)} s^{}_\rho}{\zeta^{}_{13}} -
\frac{c^{}_{(\delta -\sigma)} s^{}_\sigma}{\zeta^{}_{23}} \right )
\frac{c^{}_{12} s^{}_{12} c^{}_{23} s^{}_{23}}{s^{}_{13}} \right ]
\; ,
\nonumber \\
\frac{{\rm d} \sigma}{{\rm d} t} & \approx &
\frac{y^2_\tau}{16\pi^2} \left [\frac{c^{}_{(\rho -\sigma)}
s^{}_{(\rho -\sigma)}}{\zeta^{}_{12}} c^2_{12} s^2_{23} + \left (
\frac{c^{}_{(\delta -\rho)} s^{}_\rho}{\zeta^{}_{13}} -
\frac{c^{}_{(\delta -\sigma)} s^{}_\sigma}{\zeta^{}_{23}} \right )
\frac{c^{}_{12} s^{}_{12} c^{}_{23} s^{}_{23}}{s^{}_{13}} \right ]
\; .
%       (13)
\end{eqnarray}
Thus the trend of $\rho$ or $\sigma$ in the RGE evolution is very
similar to that of $\delta$.

\item       The Jarlskog parameter is sensitive to both $\delta$
and $\theta_{13}$. Given the initial condition $\theta_{13} =
1^\circ$, $\delta = 90^\circ$ at $\Lambda_{\rm EW}$ can be
achieved from $\delta = 0^\circ$ at $\Lambda_{\rm SS}$. It turns
out that ${\cal J} \approx 0.26\%$ can be radiatively generated in
this specific case. When $\theta_{13} = 5^\circ$ is input at
$\Lambda_{\rm SS}$, one may arrive at ${\cal J} \approx 1\%$ at
$\Lambda_{\rm EW}$, a value which might be detectable in the
future long-baseline neutrino oscillation experiments.

\item       The off-diagonal asymmetries ${\cal A}^{}_{\rm L}$ and
${\cal A}^{}_{\rm R}$ are both non-vanishing in our numerical
example. While ${\cal A}^{}_{\rm L}$ is sensitive to the input of
$\theta_{13}$, ${\cal A}^{}_{\rm R}$ is not. The reason for the
latter feature is simply that ${\cal A}^{}_{\rm R} = c^2_{13}
(s^2_{12} - s^2_{23})$ holds.
\end{itemize}
It is worth mentioning that the unitarity of $V$ requires that its
six unitarity triangles in the complex plane have the same area,
amounting to ${\cal J}/2$ \cite{FX00}. Thus the radiative
generation of $\delta$ or ${\cal J}$ geometrically implies the
radiative generation of every leptonic unitarity triangle; i.e.,
three overlapped lines (sides) at $\Lambda_{\rm SS}$ can evolve
into a triangle at $\Lambda_{\rm EW}$.

Now let us examine whether the radiative generation of $\delta =
90^\circ$ at $\Lambda_{\rm EW}$ can be achieved from other initial
values of $\rho$ and $\sigma$, when $\theta_{13} = 1^\circ$ holds
and $m^{}_1$, $\Delta m^2_{21}$, $\Delta m^2_{31}$, $\theta_{12}$
and $\theta_{23}$ take the same input values as before at
$\Lambda_{\rm SS}$ (see Table I). We find out two new numerical
examples with $(\rho, ~\sigma) = (0^\circ, -59.9^\circ)$ and
$(10^\circ, -57^\circ)$, respectively, as shown in FIG. 2. Note
that both $\delta$ and $\rho$ (or $\sigma$) can be radiatively
generated from $\sigma \neq 0^\circ$ (or $\rho \neq 0^\circ$) at
the seesaw scale
%%%%%%%%%%%%%%%%%%%%%%%%%%
\footnote{The possibility to simultaneously generate $\rho$ and
$\sigma$ from $\delta \neq 0^\circ$ at $\Lambda_{\rm SS}$ via the
RGE evolution is in general expected to be strongly suppressed,
because the leading terms of ${\rm d}\rho/{\rm d}t$ and ${\rm
d}\sigma/{\rm d}t$ in Eq. (7) vanish for $\rho = \sigma = 0^\circ$
at $\Lambda_{\rm SS}$. Only in the $\theta_{13} \rightarrow 0$
limit, the running effects of three CP-violating phases could
become significant \cite{Mei}.}.
%%%%%%%%%%%%%%%%%%%%%%%%%%
Our results hint at the existence of strong parameter degeneracy
in obtaining $\delta = 90^\circ$ at $\Lambda_{\rm EW}$ from
$\delta = 0^\circ$ at $\Lambda_{\rm SS}$. To resolve this problem
is certainly a big challenge in model building, unless two
Majorana CP-violating phases could separately be measured at low
energies.

Finally we demonstrate that it is also possible to radiatively
generate $\delta = 90^\circ$ at $\Lambda_{\rm EW}$ from $\delta =
0^\circ$ at $\Lambda_{\rm SS}$ in the $\Delta m^2_{31} < 0$ case.
Our numerical results are shown in FIG. 3, where the initial
values of relevant parameters are listed in Table II. One can see
that the one-loop RGE running behaviors of three CP-violating
phases in FIG. 3 are very similar to those in FIG. 1, although the
initial conditions in these two cases are quite different. The
reason for this similarity has actually been reflected by Eq.
(13), where the sign of $\Delta m^2_{31}$ does not play a role in
the leading-order approximation.

\subsection{Radiative Generation of $\rho=90^\circ$}

We continue to take a look at the one-loop RGE evolution of $\rho$
in the $m^{}_1 < m^{}_2 < m^{}_3$ case (i.e., $\Delta m^2_{31} >
0$). The input and output values of relevant physical parameters
are listed in Table III. One can see that there is no difficulty
to radiatively generate $\rho = 90^\circ$ at $\Lambda_{\rm EW}$
from $\rho = 0^\circ$ at $\Lambda_{\rm SS}$, provided $\theta_{13}
= 1^\circ$, $\delta = 0^\circ$ and $\sigma = -67.7^\circ$ are
input. Allowing the initial value of $\theta_{13}$ to change to
$3^\circ$ or $5^\circ$ but fixing the input values of the other
quantities, we find that only $\rho = 17.6^\circ$ or $\rho =
12.1^\circ$ can be obtained at $\Lambda_{\rm EW}$. Again the
results of $m^{}_1$, $\Delta m^2_{31}$, $\theta_{12}$ and
$\theta_{23}$ at $\Lambda_{\rm EW}$ are quite stable against the
change of $\theta_{13}$ from $1^\circ$ to $5^\circ$ at
$\Lambda_{\rm SS}$, but the result of $\Delta m^2_{21}$ becomes
smaller.

The RGE running behaviors of three CP-violating phases $(\delta,
\rho, \sigma)$ and three rephasing-invariant parameters $({\cal
J}, {\cal A}^{}_{\rm L}, {\cal A}^{}_{\rm R})$ are explicitly
shown in FIG. 4. Comparing this figure with FIG. 1, one may
observe much similarity between them. There are two reasons for
this similarity: first, the initial conditions in these two cases
are not very different; second, the evolution of $\delta$, $\rho$
and $\sigma$ is dominated by Eq. (13) in both cases.

\subsection{Radiative Generation of $\sigma=90^\circ$}

Finally let us look at the RGE evolution of $\sigma$ in the
$m^{}_1 < m^{}_2 < m^{}_3$ case (i.e., $\Delta m^2_{31} > 0$). The
input and output values of relevant physical parameters are listed
in Table IV. One can see that there is no difficulty to
radiatively generate $\sigma = 90^\circ$ at $\Lambda_{\rm EW}$
from $\sigma = 0^\circ$ at $\Lambda_{\rm SS}$, if $\theta_{13} =
1^\circ$, $\delta = 119.7^\circ$ and $\rho = 60.8^\circ$ are
input. Allowing the initial value of $\theta_{13}$ to change to
$3^\circ$ or $5^\circ$ but fixing the input values of the other
quantities, we find that only $\sigma = 34.4^\circ$ or $\sigma =
29.8^\circ$ can be obtained at $\Lambda_{\rm EW}$. The results of
$m^{}_1$, $\Delta m^2_{31}$, $\theta_{12}$ and $\theta_{23}$ at
$\Lambda_{\rm EW}$ are very stable against the change of
$\theta_{13}$ from $1^\circ$ to $5^\circ$ at $\Lambda_{\rm SS}$,
but the result of $\Delta m^2_{21}$ becomes larger and less
favored.

The RGE running behaviors of $\delta$, $\rho$ and $\sigma$ are
shown in FIG. 5, in which those of ${\cal J}$, ${\cal A}^{}_{\rm
L}$ and ${\cal A}^{}_{\rm R}$ are also illustrated. Again the
evolution of three CP-violating phases is dominated by Eq. (13).
Because $\delta$ evolves from the second quadrant to the third one
during the RGE running, a flip of the sign of ${\cal J}$ appears
around $\delta = 180^\circ$. The signs of ${\cal A}^{}_{\rm L}$
and ${\cal A}^{}_{\rm R}$ keep unchanged from $\Lambda_{\rm SS}$
to $\Lambda_{\rm EW}$, implying that the geometrical structure of
$V$ does not change in a significant way.

In all the cases discussed above, $\langle m\rangle_{ee} \sim
m^{}_1$ holds as a consequence of the approximate mass degeneracy
of three light neutrinos. Indeed, the smallness of $\theta_{13}$
allows us to obtain an approximate expression of $\langle
m\rangle_{ee}$:
\begin{equation}
\langle m\rangle_{ee} \approx m^{}_1 \sqrt{1 - \sin^2 2\theta_{12}
\sin^2 (\rho - \sigma)} \in \left [m^{}_1 \cos 2\theta_{12}, ~
m^{}_1 \right ] \; .
%       (14)
\end{equation}
Given $m^{}_1 \approx 0.2$ eV and $\theta_{12} \approx 33^\circ$
at low-energy scales, $\langle m\rangle_{ee}$ turns out to lie in
the range $0.08 ~ {\rm eV} \leq \langle m\rangle_{ee} \leq 0.2 ~
{\rm eV}$, a result consistent with the present experimental upper
limit $\langle m\rangle_{ee} < 0.38$ eV \cite{Fit}.

\section{Summary}

Taking account of a very useful parametrization of the $3\times 3$
MNS matrix $V$, we have derived the one-loop RGEs for its three
mixing angles $(\theta_{12}, \theta_{23}, \theta_{13})$, three
CP-violating phases $(\delta, \rho, \sigma)$ and three
rephasing-invariant quantities $({\cal J}, {\cal A}^{}_{\rm L},
{\cal A}^{}_{\rm R})$. Particular attention has been paid to the
radiative generation of leptonic CP violation, because the Dirac
phase $\delta$ and the Majorana phases $\rho$ and $\sigma$ are
entangled with one another in the RGE evolution from the seesaw
scale $\Lambda_{\rm SS}$ to the electroweak scale $\Lambda_{\rm
EW}$. We have shown that $\delta =90^\circ$ at $\Lambda_{\rm EW}$
can be radiatively generated from $\delta =0^\circ$ at
$\Lambda_{\rm SS}$ in the minimal supersymmetric standard model,
provided the light neutrino masses are nearly degenerate and the
mixing angle $\theta_{13}$ is of ${\cal O}(1^\circ)$ or smaller.
As for $\rho$ and $\sigma$, it is also possible to radiatively
generate $\rho =90^\circ$ or $\sigma = 90^\circ$ at $\Lambda_{\rm
EW}$ from $\rho =0^\circ$ or $\sigma = 0^\circ$ at $\Lambda_{\rm
SS}$. An interesting feature of our numerical analysis is that the
one-loop RGE running behaviors of three CP-violating phases are
quite similar in the chosen parameter space, no matter whether the
sign of the neutrino mass-squared difference $\Delta m^2_{31}$ is
positive or negative. As an important by-product, the geometrical
structure of $V$ (i.e., its off-diagonal asymmetries and the area
of its unitarity triangles in the complex plane) against radiative
corrections has also been discussed.

At this point, it is worthwhile to summarize some remarkable
differences between our analytical and numerical results and those
obtained in Refs. \cite{RGE2,RGE3}:

(1) The phase convention of $V$ used in Refs. \cite{RGE2,RGE3}
leads to the apparent dependence of $\langle m\rangle_{ee}$ (the
effective mass of the neutrinoless double-beta decay) on $\delta$,
thus it is ill to refer to $\delta$ as the ``Dirac" CP-violating
phase. Such an ambiguity has been avoided in our phase convention
of $V$, which forbids $\delta$ to enter the expression of $\langle
m\rangle_{ee}$. Our Majorana phases ($\rho$ and $\sigma$) are
related to the ones defined by Casas {\it et al} \cite{RGE2}
($\phi$ and $\phi'$) or by Antusch {\it et al} \cite{RGE3}
($\varphi^{}_1$ and $\varphi^{}_2$) as follows: $\varphi^{}_1 =
\phi = 2 \left (\delta - \rho \right )$ and $\varphi^{}_2 = \phi'
= 2 \left (\delta - \sigma \right )$. On the other hand, our Dirac
phase $\delta$ is indistinguishable from their phase parameter
$\delta$ in the description of leptonic CP violation in neutrino
oscillations.

(2) Our RGEs for three mixing angles and three CP-violating phases
are apparently different from those given in Refs.
\cite{RGE2,RGE3}, just because we have taken a different and more
instructive phase convention for $V$. In particular, the leading
terms of ${\rm d}\rho/{\rm d}t$ and ${\rm d}\sigma/{\rm d}t$
(i.e., the terms proportional to $1/\zeta^{}_{12}$ and
$1/s^{}_{13}$) are very similar to those of ${\rm d}\delta/{\rm
d}t$, as one can see from Eq. (7) or Eq. (13). This result implies
that three CP-violating phases have similar RGE running behaviors
in our phase convention. In contrast, the RGE running effect on
$\varphi^{}_1$ (or $\phi$) and $\varphi^{}_2$ (or $\phi'$) is much
weaker than that on $\delta$ in Refs. \cite{RGE2,RGE3}.

(3) The one-loop RGEs for ${\cal J}$, ${\cal A}_{\rm L}$ and
${\cal A}_{\rm R}$ given in Eqs. (10) and (11) are new results,
although they can be derived from Eqs. (6) and (7) in a
straightforward way. These three quantities are actually
rephasing-invariant or parametrization-independent. Hence we can
get some more generic feeling about radiative corrections to the
leptonic CP-violating effect and the geometrical structure of $V$
from the RGE evolution of ${\cal J}$, ${\cal A}_{\rm L}$ and
${\cal A}_{\rm R}$. This kind of study was not done in Refs.
\cite{RGE2,RGE3} or elsewhere.

(4) Different from those previous works, we have concentrated on
the novel possibilities to radiatively generate $\delta =
90^\circ$, $\rho = 90^\circ$ or $\sigma = 90^\circ$ at
$\Lambda_{\rm EW}$ from $\delta = 0^\circ$, $\rho =0^\circ$ or
$\sigma =0^\circ$ at $\Lambda_{\rm SS}$ in our numerical
exercises. Because of ${\cal J} \propto \sin\delta$, $\delta \sim
90^\circ$ at low energies is a necessary condition to achieve
sufficiently large CP- and $T$-violating effects in neutrino
oscillations. On the other hand, $\rho  \sim 90^\circ$ or $\sigma
\sim 90^\circ$ at low energies may result in a kind of large
cancellation in $\langle m\rangle_{ee}$, implying the possible
suppression of this unique experimental observable to identify the
Majorana nature of massive neutrinos.

It is also worth remarking that our analysis is essentially
independent of the specific textures of lepton Yukawa coupling
matrices, thus it can be applied to the concrete work of model
building. Since the elegant thermal leptogenesis mechanism is
usually expected to work at the seesaw scale, a study of its
consequences at low-energy scales is available by means of the
one-loop RGEs that we have obtained. In other words, the RGEs may
serve as a useful bridge to establish a kind of connection between
the phenomena of CP violation at low- and high-energy scales. It
would be extremely interesting, in our opinion, if the phase
parameter governing the strength of CP violation in a
long-baseline neutrino oscillation experiment could be radiatively
generated from those CP-violating phases which control the
matter-antimatter asymmetry of our universe at the seesaw scale.

\vspace{0.5cm}

One of us (Z.Z.X.) is grateful to S. Zhou for very helpful
discussions. This work was supported in part by the National
Nature Science Foundation of China.

%\newpage

\newpage

%%%%%%%%%%%%%%%%%%%%% Table 1 %%%%%%%%%%%%%%%%%%%%%%%%
\begin{table}
\caption{Radiative generation of $\delta = 90^\circ$ at the
electroweak scale  $\Lambda_{\rm EW}$ from $\delta =0^\circ$ at
the seesaw scale $\Lambda_{\rm SS}$ in the MSSM with $\tan\beta =
10$ and $\Delta m^2_{31} >0$.} \vspace{0.2cm}
\begin{center}
\begin{tabular}{c|l|ccc}
Parameter & Input $\left(\Lambda_{\rm SS}^{}\right)$ ~~~~~~ &
\multicolumn{3}{c}{Output $( \Lambda_{\rm EW} )$} \\ & &
$\theta_{13}
= 1^\circ$ & $\theta_{13} = 3^\circ$ & $\theta_{13} = 5^\circ$ \\
\hline
$m^{}_1 ({\rm eV} )$ & 0.241 & 0.20 & 0.20 & 0.20 \\
%-----------------------------
$\Delta m^2_{21} ( 10^{-5} ~{\rm eV}^2 )$ & 20.4 & 7.79 & 7.17 & 6.56 \\
%-----------------------------
$\Delta m^2_{31} ( 10^{-3} ~{\rm eV}^2 )$ & 3.32 & 2.20 & 2.20 & 2.20 \\
\hline
%=============================
$\theta_{12}$ & $24.1^\circ$ & $33.0^\circ$ & $33.0^\circ$ & $33.1^\circ$ \\
%-----------------------------
$\theta_{23}$ & $43.9^\circ$ & $45.1^\circ$ & $45.0^\circ$ & $45.0^\circ$ \\
%-----------------------------
$\theta_{13}$ & $1^\circ/3^\circ/5^\circ$ & $0.65^\circ$ & $2.46^\circ$ & $4.52^\circ$ \\
\hline
%-----------------------------
$\delta$ & $0^\circ$ & $90.0^\circ$ & $41.8^\circ$ & $35.8^\circ$ \\
%-----------------------------
$\rho$ & $4.0^\circ$ & $72.2^\circ$ & $23.8^\circ$ & $17.6^\circ$ \\
%-----------------------------
$\sigma$ & $-57.5^\circ$ & $26.3^\circ$ & $-22.0^\circ$ & $-28.1^\circ$ \\
\end{tabular}
\end{center}
\end{table}
%%%%%%%%%%%%%%%%%%%%%%%%%%%%%%%%%%%%%%%%%%%%%%%%%%%%%%%%%%%%%%%%%%%%%

%%%%%%%%%%%%%%%%%%%%% Table 2 %%%%%%%%%%%%%%%%%%%%%%%%
\begin{table}
\caption{Radiative generation of $\delta = 90^\circ$ at the
electroweak scale  $\Lambda_{\rm EW}$ from $\delta =0^\circ$ at
the seesaw scale $\Lambda_{\rm SS}$ in the MSSM with $\tan\beta =
10$ and $\Delta m^2_{31} < 0$.} \vspace{0.2cm}
\begin{center}
\begin{tabular}{c|l|ccc}
Parameter & Input $\left(\Lambda_{\rm SS}^{}\right)$ ~~~~~ &
\multicolumn{3}{c}{Output $( \Lambda_{\rm EW}
)$} \\
& & $\theta_{13}
= 1^\circ$ & $\theta_{13} = 3^\circ$ & $\theta_{13} = 5^\circ$ \\
\hline
$m^{}_1 ({\rm eV} )$ & 0.241 & 0.20 & 0.20 & 0.20 \\
%-----------------------------
$\Delta m^2_{21} ( 10^{-5} ~{\rm eV}^2 )$ & 20.8 & 7.96 & 7.37 & 6.79 \\
%-----------------------------
$\Delta m^2_{31} ( 10^{-3} ~{\rm eV}^2 )$ & -3.08 & -2.20 & -2.21 & -2.21 \\
\hline
%=============================
$\theta_{12}$ & $25.1^\circ$ & $33.3^\circ$ & $33.5^\circ$ & $33.7^\circ$ \\
%-----------------------------
$\theta_{23}$ & $46.7^\circ$ & $45.2^\circ$ & $45.2^\circ$ & $45.2^\circ$ \\
%-----------------------------
$\theta_{13}$ & $1^\circ/3^\circ/5^\circ$ & $0.68^\circ$ & $2.38^\circ$ & $4.34^\circ$ \\
\hline
%-----------------------------
$\delta$ & $0^\circ$ & $90.0^\circ$ & $43.1^\circ$ & $36.8^\circ$ \\
%-----------------------------
$\rho$ & $-81.9^\circ$ & $-13.6^\circ$ & $-60.7^\circ$ & $-67.2^\circ$ \\
%-----------------------------
$\sigma$ & $34.3^\circ$ & $117.7^\circ$ & $70.6^\circ$ & $64.1^\circ$ \\
\end{tabular}
\end{center}
\end{table}
%%%%%%%%%%%%%%%%%%%%%%%%%%%%%%%%%%%%%%%%%%%%%%%%%%%%%%%%%%%%%%%%%%%%%

%%%%%%%%%%%%%%%%%%%%% Table 3 %%%%%%%%%%%%%%%%%%%%%%%%
\begin{table}
\caption{Radiative generation of $\rho = 90^\circ$ at the
electroweak scale  $\Lambda_{\rm EW}$ from $\rho =0^\circ$ at the
seesaw scale $\Lambda_{\rm SS}$ in the MSSM with $\tan\beta = 10$
and $\Delta m^2_{31} >0$.} \vspace{0.2cm}
\begin{center}
\begin{tabular}{c|l|ccc}
Parameter & Input $\left(\Lambda_{\rm SS}^{}\right)$ ~~~~~ &
\multicolumn{3}{c}{Output $( \Lambda_{\rm EW}
)$} \\
& & $\theta_{13}
= 1^\circ$ & $\theta_{13} = 3^\circ$ & $\theta_{13} = 5^\circ$ \\
\hline
$m^{}_1 ({\rm eV} )$ & 0.241 & 0.20 & 0.20 & 0.20 \\
%-----------------------------
$\Delta m^2_{21} ( 10^{-5} ~{\rm eV}^2 )$ & 20.4 & 8.54 & 7.90 & 7.27 \\
%-----------------------------
$\Delta m^2_{31} ( 10^{-3} ~{\rm eV}^2 )$ & 3.32 & 2.21 & 2.20 & 2.20 \\
\hline
%=============================
$\theta_{12}$ & $27.6^\circ$ & $33.1^\circ$ & $33.2^\circ$ & $33.3^\circ$ \\
%-----------------------------
$\theta_{23}$ & $43.9^\circ$ & $44.8^\circ$ & $44.8^\circ$ & $44.8^\circ$ \\
%-----------------------------
$\theta_{13}$ & $1^\circ/3^\circ/5^\circ$ & $0.43^\circ$ & $2.17^\circ$ & $4.24^\circ$ \\
\hline
%-----------------------------
$\delta$ & $0^\circ$ & $107.6^\circ$ & $35.4^\circ$ & $30.2^\circ$ \\
%-----------------------------
$\rho$ & $0^\circ$ & $90.2^\circ$ & $17.6^\circ$ & $12.1^\circ$ \\
%-----------------------------
$\sigma$ & $-67.7^\circ$ & $34.1^\circ$ & $-38.3^\circ$ & $-44.6^\circ$ \\
\end{tabular}
\end{center}
\end{table}
%%%%%%%%%%%%%%%%%%%%%%%%%%%%%%%%%%%%%%%%%%%%%%%%%%%%%%%%%%%%%%%%%%%%%

%%%%%%%%%%%%%%%%%%%%% Table 4 %%%%%%%%%%%%%%%%%%%%%%%%
\begin{table}
\caption{Radiative generation of $\sigma = 90^\circ$ at the
electroweak scale  $\Lambda_{\rm EW}$ from $\sigma =0^\circ$ at
the seesaw scale $\Lambda_{\rm SS}$ in the MSSM with $\tan\beta =
10$ and $\Delta m^2_{31} >0$.} \vspace{0.2cm}
\begin{center}
\begin{tabular}{c|l|ccc}
Parameter & Input $\left(\Lambda_{\rm SS}^{}\right)$ ~~~~~ &
\multicolumn{3}{c}{Output $( \Lambda_{\rm EW}
)$} \\
& & $\theta_{13}
= 1^\circ$ & $\theta_{13} = 3^\circ$ & $\theta_{13} = 5^\circ$ \\
\hline
$m^{}_1 ({\rm eV} )$ & 0.241 & 0.20 & 0.20 & 0.20 \\
%-----------------------------
$\Delta m^2_{21} ( 10^{-5} ~{\rm eV}^2 )$ & 20.3 & 8.24 & 8.85 & 9.50 \\
%-----------------------------
$\Delta m^2_{31} ( 10^{-3} ~{\rm eV}^2 )$ & 3.32 & 2.21 & 2.21 & 2.21 \\
\hline
%=============================
$\theta_{12}$ & $24.1^\circ$ & $33.2^\circ$ & $34.2^\circ$ & $35.2^\circ$ \\
%-----------------------------
$\theta_{23}$ & $43.9^\circ$ & $44.6^\circ$ & $44.6^\circ$ & $44.6^\circ$ \\
%-----------------------------
$\theta_{13}$ & $1^\circ/3^\circ/5^\circ$ & $0.51^\circ$ & $2.27^\circ$ & $4.29^\circ$ \\
\hline
%-----------------------------
$\delta$ & $119.7^\circ$ & $216.0^\circ$ & $161.0^\circ$ & $157.1^\circ$ \\
%-----------------------------
$\rho$ & $60.8^\circ$ & $135.3^\circ$ & $78.8^\circ$ & $73.6^\circ$ \\
%-----------------------------
$\sigma$ & $0^\circ$ & $90.0^\circ$ & $34.4^\circ$ & $29.8^\circ$ \\
\end{tabular}
\end{center}
\end{table}
%%%%%%%%%%%%%%%%%%%%%%%%%%%%%%%%%%%%%%%%%%%%%%%%%%%%%%%%%%%%%%%%%%%%%

\newpage

%%%%%%%%%%%%%%%%%%%%% Fig. 1 %%%%%%%%%%%%%%%%%%%
\begin{figure}
\begin{center}
\vspace{-1cm}
\includegraphics[width=15.5cm,height=22.5cm]{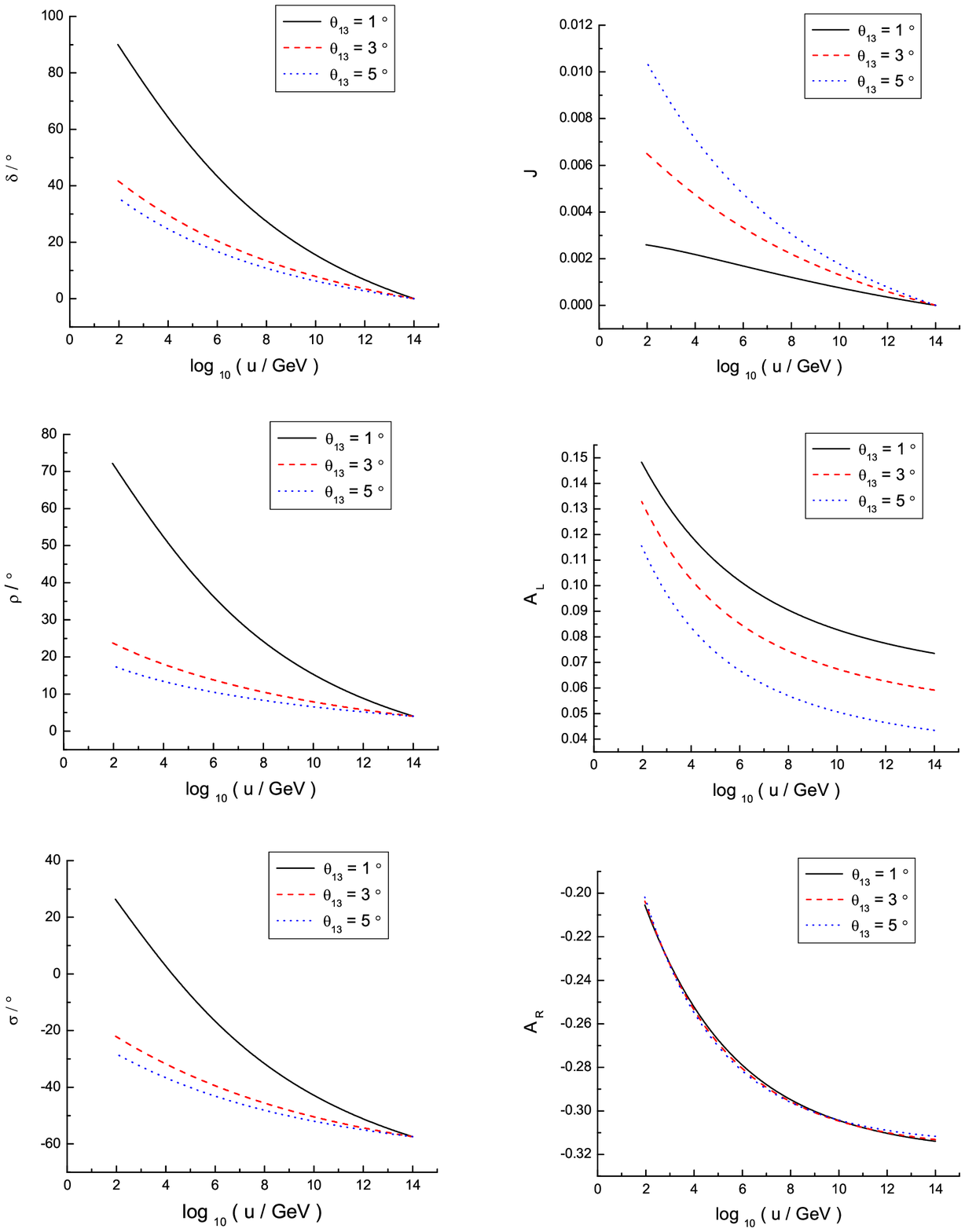}
\vspace{-3.cm}
\caption{The RGE running behaviors of three
CP-violating phases and three rephasing-invariant quantities of
$V$ from $\Lambda_{\rm SS}$ to $\Lambda_{\rm EW}$ in the MSSM,
where the input values of relevant parameters can be found from
Table I.}
\end{center}
\end{figure}
%%%%%%%%%%%%%%%%%%%%%%%%%%%%%%%%%%%%%%%%%%%%%%%

\newpage

%%%%%%%%%%%%%%%%%%%%% Fig. 2 %%%%%%%%%%%%%%%%%%%
\begin{figure}
\begin{center}
\vspace{-1cm}
\includegraphics[width=15.5cm,height=22.5cm]{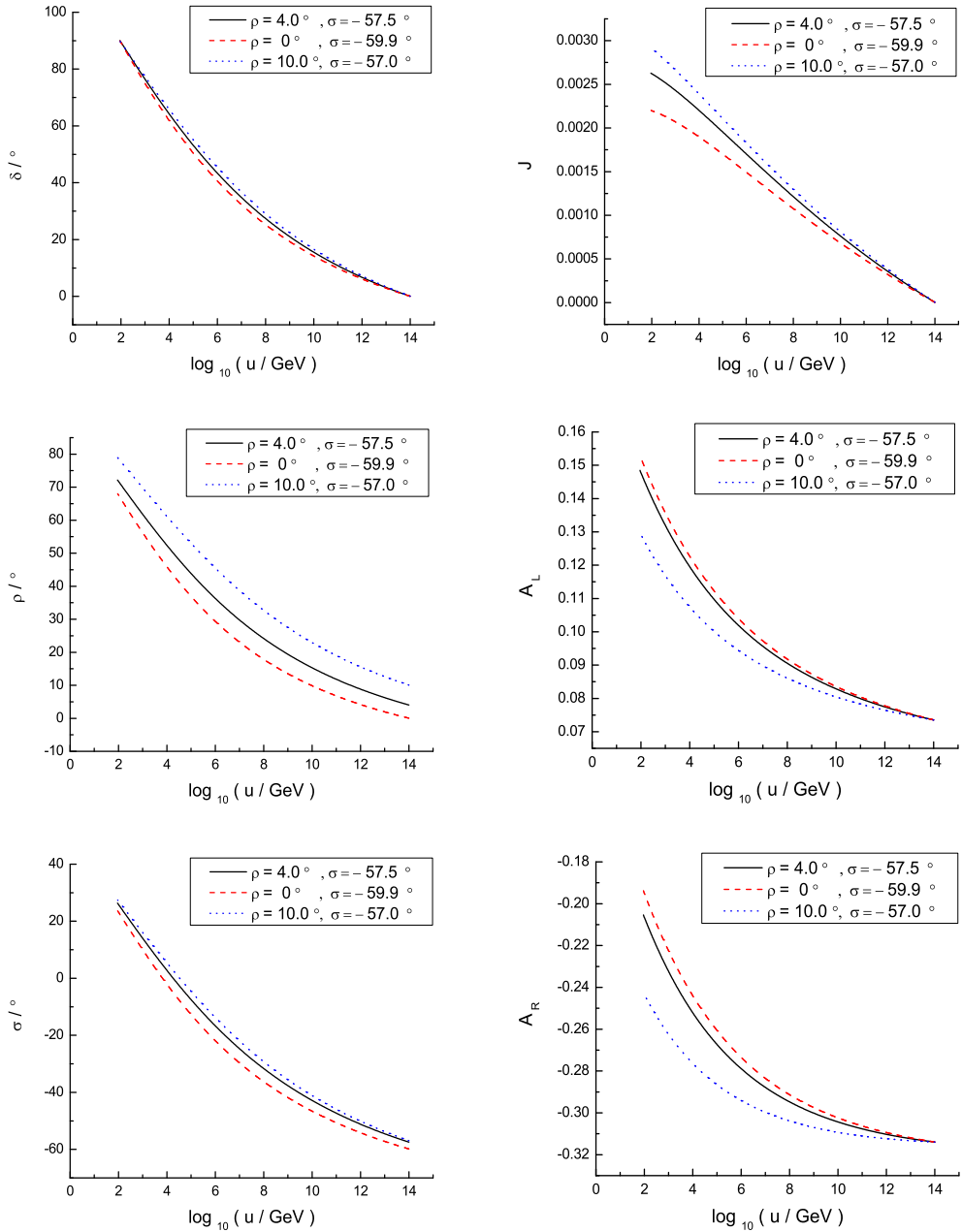}
\vspace{-3cm}
\caption{The radiative generation of $\delta =
90^\circ$ at $\Lambda_{\rm EW}$ from $\theta_{13} = 1^\circ$ and
different values of $\rho$ and $\sigma$ at $\Lambda_{\rm SS}$. The
input values of other parameters can be found from Table I.}
\end{center}
\end{figure}
%%%%%%%%%%%%%%%%%%%%%%%%%%%%%%%%%%%%%%%%%%%%%%%

\newpage

%%%%%%%%%%%%%%%%%%%%% Fig. 3 %%%%%%%%%%%%%%%%%%%
\begin{figure}
\begin{center}
\vspace{-1cm}
\includegraphics[width=15.5cm,height=22.5cm]{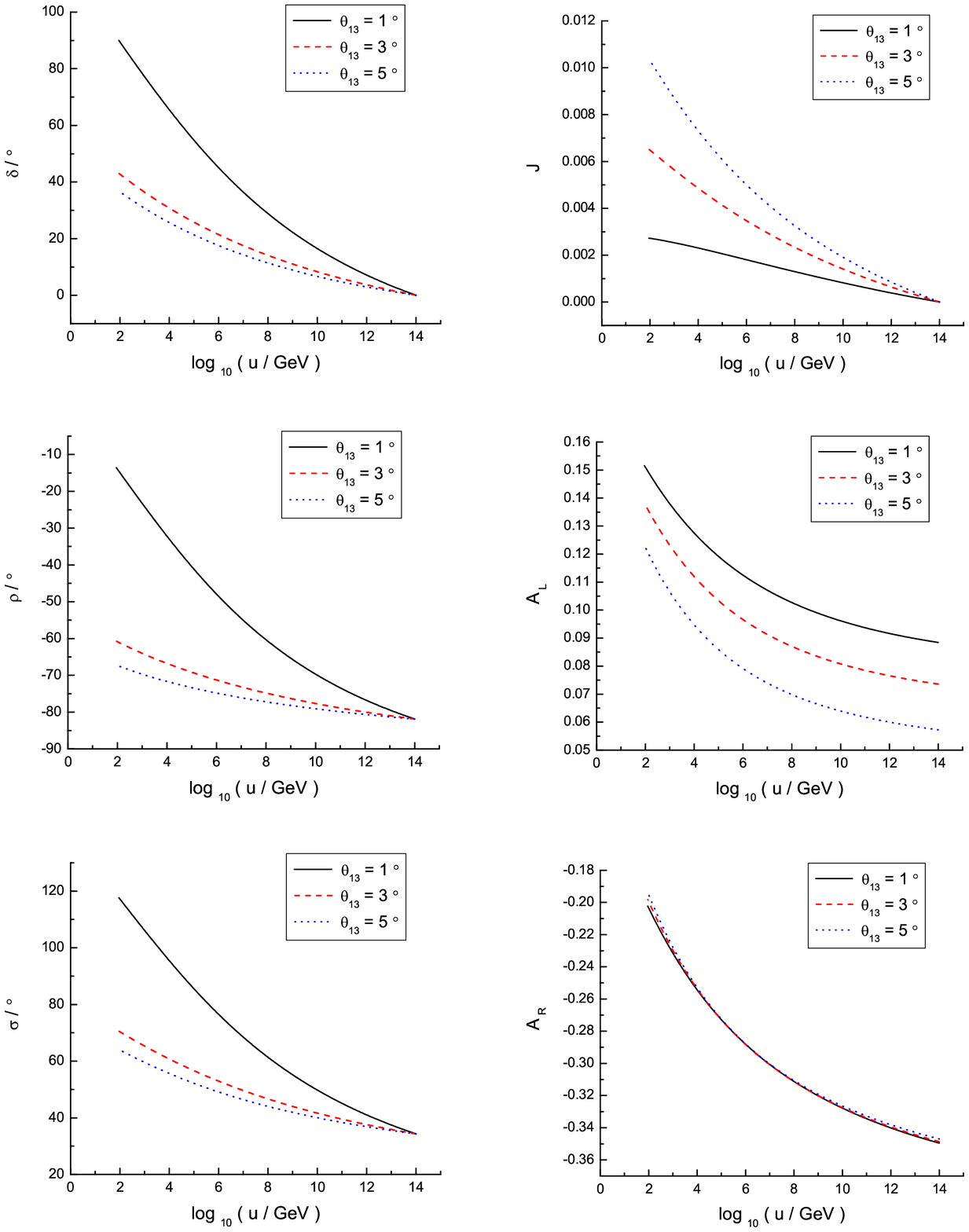}
\vspace{-3cm}
\caption{The RGE running behaviors of three
CP-violating phases and three rephasing-invariant quantities of
$V$ from $\Lambda_{\rm SS}$ to $\Lambda_{\rm EW}$ in the MSSM,
where the input values of relevant parameters can be found from
Table II.}
\end{center}
\end{figure}
%%%%%%%%%%%%%%%%%%%%%%%%%%%%%%%%%%%%%%%%%%%%%%%

\newpage

%%%%%%%%%%%%%%%%%%%%% Fig. 4 %%%%%%%%%%%%%%%%%%%
\begin{figure}
\begin{center}
\vspace{-1cm}
\includegraphics[width=15.5cm,height=22.5cm]{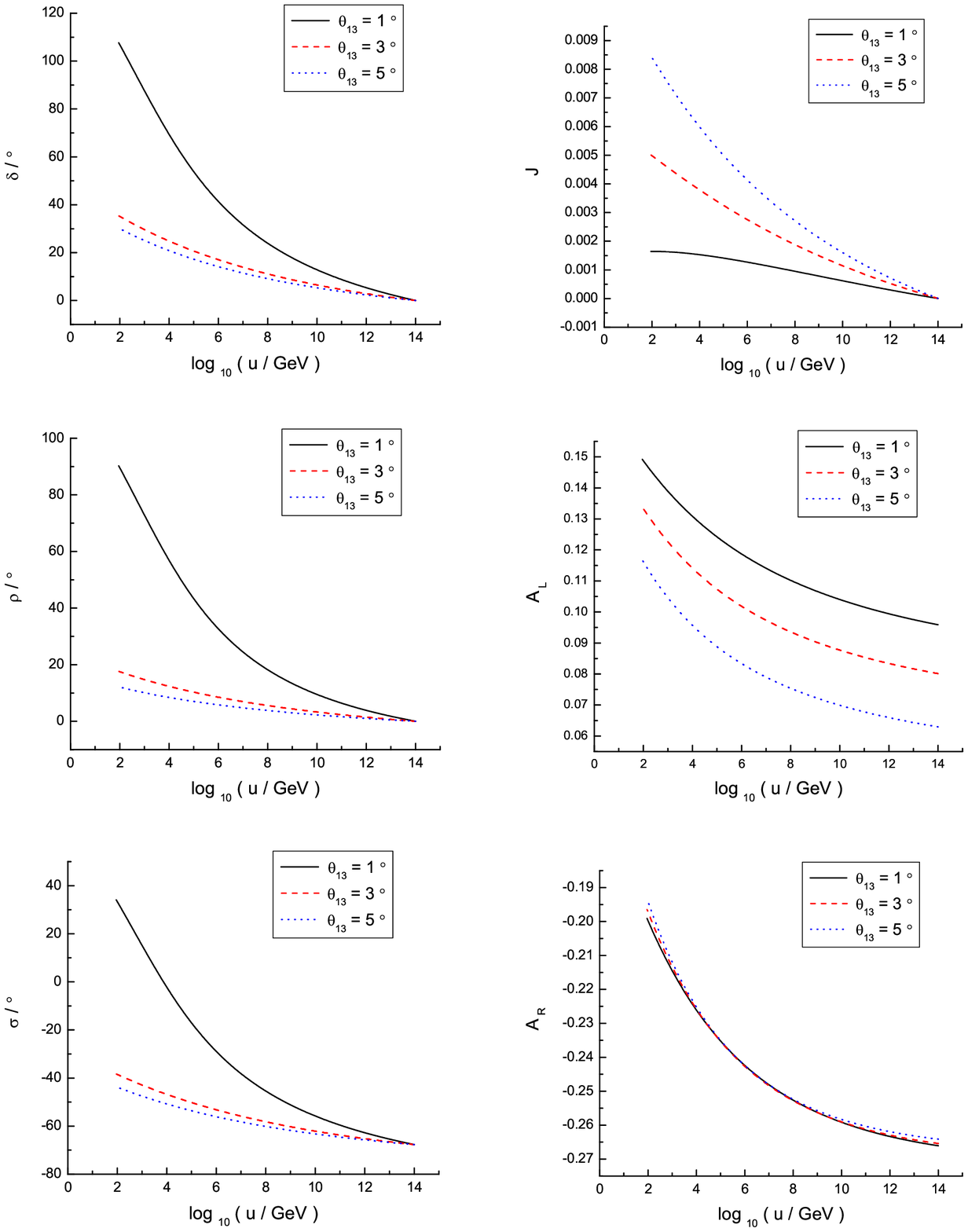}
\vspace{-3.cm}
\caption{The RGE running behaviors of three
CP-violating phases and three rephasing-invariant quantities of
$V$ from $\Lambda_{\rm SS}$ to $\Lambda_{\rm EW}$ in the MSSM,
where the input values of relevant parameters can be found from
Table III.}
\end{center}
\end{figure}
%%%%%%%%%%%%%%%%%%%%%%%%%%%%%%%%%%%%%%%%%%%%%%%

\newpage

%%%%%%%%%%%%%%%%%%%%% Fig. 5 %%%%%%%%%%%%%%%%%%%
\begin{figure}
\begin{center}
\vspace{-1cm}
\includegraphics[width=15.5cm,height=22.5cm]{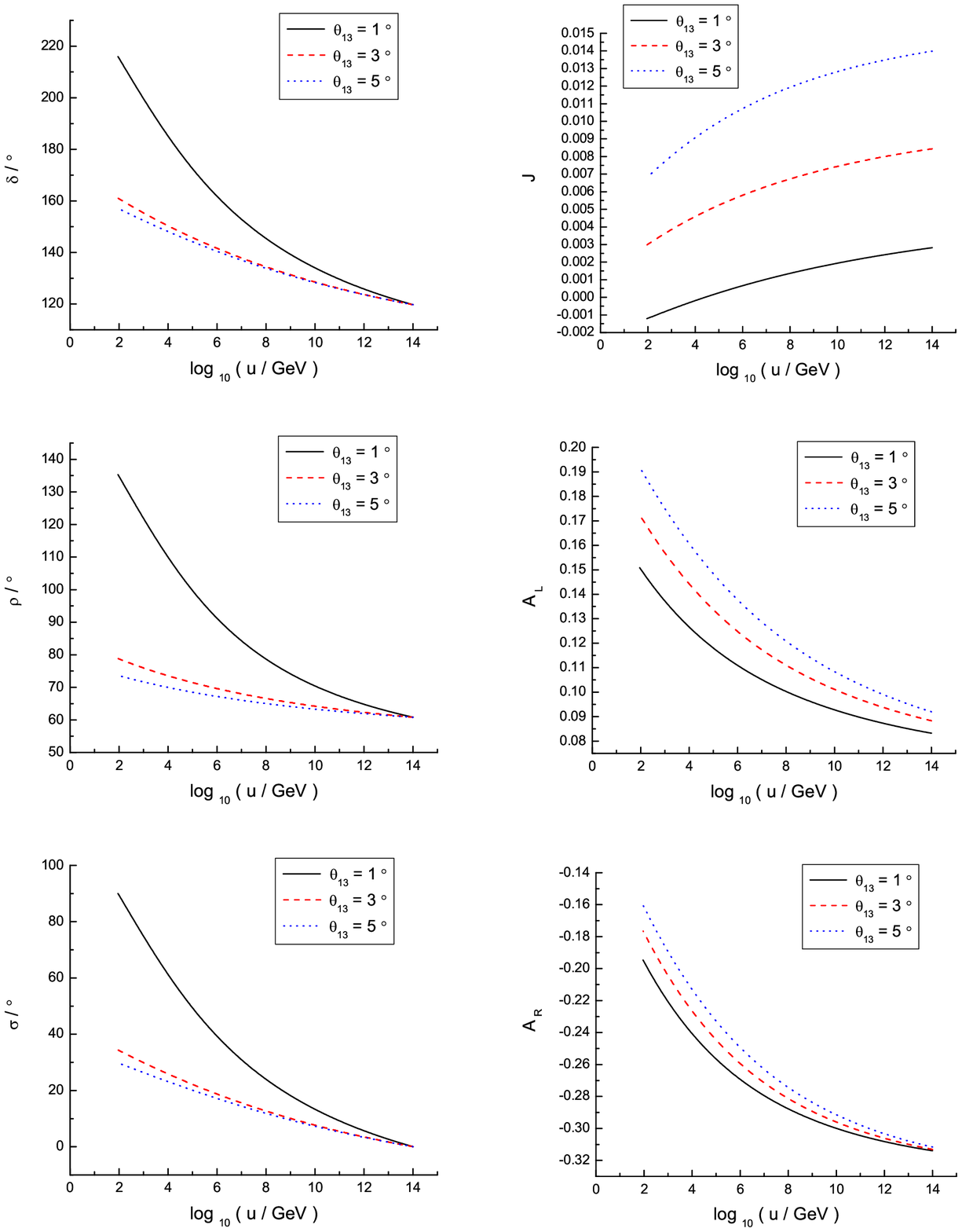}
\vspace{-3.cm}
\caption{The RGE running behaviors of three
CP-violating phases and three rephasing-invariant quantities of
$V$ from $\Lambda_{\rm SS}$ to $\Lambda_{\rm EW}$ in the MSSM,
where the input values of relevant parameters can be found from
Table IV.}
\end{center}
\end{figure}
%%%%%%%%%%%%%%%%%%%%%%%%%%%%%%%%%%%%%%%%%%%%%%%

\end{document}